\documentclass[11pt,paper]{JHEP3}
\date{March 2008}
\usepackage{epsfig}
\usepackage{latexsym}
\usepackage{amssymb}
\usepackage{booktabs}
\newcommand{\be}{\begin{equation}}
\newcommand{\ee}{\end{equation}}
\newcommand{\ba}{\begin{eqnarray}}
\newcommand{\ea}{\end{eqnarray}}
\newcommand{\bi}{\begin{itemize}}
\newcommand{\ei}{\end{itemize}}

\newcommand{\nn}{\nonumber \\}

\newcommand{\quarter}{{\textstyle\frac{1}{4}}}

\newcommand{\<}{\langle}
\renewcommand{\>}{\rangle}
\newcommand{\eq}{Eq.~}

\newcommand{\la}{\label}

\newcommand{\txts}{\textstyle}

\newcommand{\K}{{\cal K}}
\newcommand{\J}{{\cal J}}
\newcommand{\I}{{\cal I}}

\title{Finite Volume Effects in Thermal Field Theory}

\author{
Harvey~B.~Meyer \\
Center for Theoretical Physics\\
Massachusetts Institute of Technology\\
Cambridge, MA 02139, U.S.A.\\
E-mail: \email{meyerh@mit.edu}
}

\preprint{MIT-CTP 4035}

\abstract{
We analyze the effect of a finite volume on the thermodynamic potentials
of a relativistic quantum field theory defined on a hypertorus
at vanishing chemical potential.
Using the symmetries of the Euclidean partition function,
we interpret the thermodynamic observables as the 
expectation value of the energy-momentum tensor in the same 
theory living on a $\beta\times L^2$ volume. In the case where the
screening correlation lengths in the thermal system are finite, 
we obtain a closed formula for the 
leading finite volume effects in terms of the smallest screening mass.
This formula can be used to estimate, and possibly correct
for, the leading finite volume effects in lattice calculations
of QCD thermodynamics.
}

\begin{document}
\section{Introduction}
The description of relativistic quantum systems at finite temperature
plays a central role in cosmology, astrophysics, plasma physics
and in the physics of heavy-ion collisions. In the latter context,
the thermodynamics of Quantum Chromodynamics (QCD) is being
studied intensively by lattice Monte-Carlo 
methods~\cite{Cheng:2007jq,Bazavov:2009zn,Aoki:2006br,Aoki:2009sc,Umeda:2008bd}
and by analytic~\cite{Kajantie:2002wa,Blaizot:2003iq} and 
semi-analytic methods~\cite{Hietanen:2008tv}. The lattice calculations 
have to control both the discretization errors and the finite volume 
effects. References~\cite{Beinlich:1995ik,Heller:1999xz,Meyer:2007fc}
address the question of how to reduce the former uncertainties. 
Here we address in some generality 
the finite-volume corrections to the energy density, entropy density 
and pressure calculated on a hypertorus of dimensions $L^3$.
At the same time, the method we follow provides a complementary
point of view on the thermodynamics of the quantum field theory.
This alternative interpretation 
might find some use in approximate analytic treatments such as 
the variational method~\cite{Kogan:1994wf,Kogan:2002yr}.

Finite size effects in  gauge theories 
have been studied before at weak coupling~\cite{Kapusta:1989cr,Engels:1981ab}.
In \cite{Gliozzi:2007jh}, an elegant calculation is presented 
that yields the finite-size effects for non-interacting gauge bosons.
For non-Abelian gauge theories at extremely high temperatures, 
these finite-volume effects are indeed expected to be 
the leading ones in the regime
$1/g\gg LT\gg 1$ ($L$ is the linear box size, $T$ is the temperature
and $g$ is the gauge coupling).
In that regime, the effects of electric and 
magnetic screening~\cite{Gross:1980br} are absent.  
However, at any finite temperature the asymptotic $LT\to\infty$
finite-size effects must be exponentially suppressed by the 
finite, non-perturbative spatial correlation length. In other words
the finite-size corrections to the thermodynamic potential is bound
to be O($e^{-cg^2TL}$), where $c$ is a number of order unity.
At a few times the deconfining temperature $T_c$, 
the screening masses are known to some extent
from four-dimensional calculations, 
both in SU(3) gauge theory~\cite{Grossman:1993wm,Datta:1998eb}
and in full QCD~\cite{Maezawa:2008kh}. 
And at much higher temperatures, the dimensional 
reduction approach allows one to predict the temperature evolution
of these correlation lengths~\cite{Hart:2000ha}. 
Choosing $L$ large compared to the 
longest of these correlation lengths should therefore ensure 
that finite-size effects are small.

In this paper we first reinterpret the finite-volume
 effects by using symmetry properties
of the Euclidean partition function and of the stress-energy 
tensor. We point out that there is a generic dynamical regime, 
where the leading finite size effects can be expressed completely
in terms of the gap in the spatial screening spectrum 
and its derivative with respect to temperature. 
This is possible because $T_{\mu\nu}$ plays the dual role of stress-energy
and energy-momentum tensor: on the one hand 
its thermal expectation value gives 
the energy density and pressure, on the other hand its 
diagonal matrix elements on an individual state yield its
energy and momentum. Since the screening
gap can be calculated on the lattice relatively easily, 
the formula we derive allows one to estimate 
the finite-size effects in practice, 
and possibly to correct for them.

In section 2 we describe the thermodynamic potentials
as expectation values of elements of the energy-momentum tensor
and the dual interpretation of these matrix elements.
In section 3 we exploit this interpretation further to
estimate the leading finite-size effects on the 
thermodynamic potentials. Numerical applications to QCD are 
presented in section 4, and we finish with some concluding remarks
(section 5).

\section{Thermodynamic observables and their dual interpretation}
We consider a relativistic theory in four space-time dimensions 
without chemical potentials in Matsubara's Euclidean formalism.
 Euclidean expectation values are denoted 
by $\<.\>$. At zero temperature and 
in infinite spatial volume, the system has a full SO($4$)
symmetry group corresponding to the Lorentz group in Minkovsky space.
As a consequence of translation invariance, 
the theory possesses a conserved, symmetric energy-momentum tensor $T_{\mu\nu}$.
We will sometimes consider separately its traceless part and its trace
$\theta$, using the notation
\be
T_{\mu\nu} = \theta_{\mu\nu}+\frac{1}{4}\theta\delta_{\mu\nu},
\qquad \theta_{\mu\mu}\equiv0.
\la{eq:Tmunu}
\ee
The conserved charges measure energy and momentum, respectively,
i.e. for a common eigenstate of these operators, we have
\ba
\int d^{3}x \,\hat T_{00}(x)|\Psi\> = E |\Psi\>\,,
\\
\int d^{3}x\, \hat T_{0k}(x)|\Psi\> = P_k |\Psi\>\,.
\ea
The partition function is $Z = \sum_n e^{-\beta E_n}$ in terms
of the eigenvalues of $\int d^{3}x \,\hat T_{00}$,
where $\beta\equiv1/T$ is the inverse temperature. 
The pressure, energy density and entropy density are obtained
from $Z$ according to 
\be
p = T\left(\frac{\partial\log Z}{\partial V}\right)_{T}+{\rm cst},\qquad
 e  = 
\frac{T^2}{V}\left(\frac{\partial\log Z}{\partial T}\right)_{V}+{\rm cst}\,,
\qquad
s = \frac{1}{V}\left(\frac{\partial (T\log Z)}{\partial T}\right)_{V}\,.
\ee
We will exclusively be considering volumes $V=L^{3}$,
with periodic boundary conditions in all four directions for bosons,
and antiperiodic boundary conditions in all four directions for fermions. 
The energy density and pressure are 
defined up to an additive constant, which we choose such that both vanish
at $\beta=L$. This is the standard choice in Monte-Carlo simulations. We have
$( e -3p)(\beta,L)=\<\theta\>_{\beta\times L^3}-\<\theta\>_{ L^4}$, 
$( e +p)(\beta,L)= \frac{4}{3}\<\theta_{00}\>_{\beta\times L^3}$ 
and, in the limit $L\to\infty$, $s=\beta( e +p)$.
We remark that one can define different operators that play the role of 
energy-momentum tensor and lead to the same conserved charges~\cite{Callan:1970ze}. 
But due to translation invariance, the Euclidean expectation values of the canonical 
and the Belinfante energy-momentum tensor 
are identical, since they differ only by a total derivative 
term~\cite{Weinberg:1995mt}.

In~\cite{Meyer:2006he},  using exact
lattice QCD sum rules~\cite{Michael:1995pv, Rothe:1995hu, Michael:1986yi}
we showed that if $|\Psi\>$ is a state of definite energy $E$ living 
in a periodic box $L_1\times L_2\times L_3$,
\ba
\<\Psi|\int d^3x\, \hat \theta_{00}(x)|\Psi\> &= &
\frac{3}{4}\left[1-\frac{1}{3}\sum_{k=1}^3 
L_k\frac{\partial}{\partial L_k}\right] E\,,
\la{eq:sr}
\\
\<\Psi|\int d^3x\, \hat \theta(x)|\Psi\> &= &
\left[1+\sum_{k=1}^3 L_k\frac{\partial}{\partial L_k}\right] E\,,
\la{eq:sr2}
\ea
and one can similarly show that 
\be
\<\Psi|{\txts\int} d^3x\,\hat \theta_{33}(x)|  \Psi\>
= -\frac{1}{4}\left[ 1 - 4 L_3 \frac{\partial}{\partial L_3}
+  \sum_{k=1}^3 L_k \frac{\partial}{\partial L_k}
\right]\, E\,.
\la{eq:th33}
\ee
The states are normalized such that $\<\Psi|\Psi\>=1$\footnote{
In the infinite volume limit, one recovers from \eq(\ref{eq:sr}--\ref{eq:th33})
the Minkovsky-space result $\<\Psi|T_{\mu\nu}|\Psi\>=P_{\mu}P_{\nu}/M$ 
for covariantly normalized one-particle states, $\<\Psi|\Psi\>=(E/M)L^3$.}.
Strictly speaking, these equations hold when taking the difference
between two states. Note however that for $|\Psi_0\>=$ 
the vacuum state in infinite volume, $\<\Psi_0|\hat\theta_{00}|\Psi_{0}\>
=\<\Psi_0|\hat\theta_{33}|\Psi_{0}\>=0$
by Euclidean symmetry. Therefore the energy appearing on the right-hand-side
of \eq(\ref{eq:sr}) and (\ref{eq:th33}) can be thought of as the energy
of the state $|\Psi\>$ relative to the infinite-volume vacuum $|\Psi_{0}\>$.
For the operator $\theta$, due to its mixing with the unit operator
one must always consider differences of matrix elements.
Finally, we expect relations \eq(\ref{eq:sr}--\ref{eq:th33})
to be true in other relativistic theories as well.

\subsection{Interchanging the coordinate axes}
Let us consider the expectation value of the operator $\theta_{00}$.
Since $\theta_{kk} \equiv -\theta_{00}$, in a $\beta\times L^3$ box 
we have  $\<\theta_{ij} \> = -\frac{\delta_{ij}}{3} \<\theta_{00}\>$.
We now want to reinterpret 
the axis labels~\footnote{The idea of interchanging the coordinate axes
 in this way is of course
not new, see for instance \cite{Arnold:1995bh}.}.
 The axis $\hat3$ will play
the role of Euclidean time (with extent $L$), while the short $\hat0$
axis assumes the role of a spatial direction (with extent $\beta$).
In the expectation values below, we always indicate the dimensions
of the lattice (in the order $\hat0,\hat1,\hat2,\hat3$).

In this new system of coordinates, the operator $\theta_{33}$ plays 
the role of the 00 component of the same tensor,
\be
\<\theta_{00}\>_{\beta\times L^3} = -3\<\theta_{33}\>_{\beta\times L^3}
=  -3\<\theta_{00}\>_{L\times (L^2\beta)}\,.
\la{eq:rot}
\ee
Next we apply formula (\ref{eq:sr}) on the energy eigenstates of the 
$\beta \times L\times L$ system.
Note that for a homogeneous state, by which we mean $E\propto L_1L_2L_3$, 
this expression vanishes. However, we will apply this on the 
lowest-energy state of the $\beta \times L\times L$ system.
We write the energy levels of that system $\tilde E_0$, $\tilde E_1$, 
$\tilde E_2$ etc. ordered by increasing energy.
We expect the energy per unit volume to have 
a finite limit when $L\to\infty$,
\be
\tilde e_0(\beta) =
 \lim_{L\to\infty}\left(\frac{\tilde E_0(\beta)  }{\beta L^2} 
-\frac{\tilde E_0(L)}{L^3}\right)\,.
\ee
The energy density $\tilde e_0(\beta)$ is thus measured 
relative to the infinite-space vacuum.
In this section we take the limit $L\to\infty$ 
in \eq(\ref{eq:rot}) and assume that  therefore
the expectation value of a local operator is equal to its
expectation value in the ground state of energy $\tilde E_0$.
A sufficient condition for this is that there should be a spectral
gap between $\tilde E_1$ and $\tilde E_0$. By combining 
\eq(\ref{eq:sr}) and \eq(\ref{eq:rot}),  we learn that 
the entropy density of the thermal system corresponds to
\be
s = \frac{4\beta }{3}\<\theta_{00}\>_{\beta\times \infty^3} 
= 
\beta ^2 \,\frac{\partial\tilde e_0(\beta)}{\partial \beta }\,.
\ee
Similarly, using \eq(\ref{eq:sr2}) one easily finds that
\be
 e -3p = 4\tilde e_0(\beta) + 
\beta \frac{\partial \tilde e_0(\beta)}{\partial\beta}
\ee
By taking a  linear combination of the last two equations, we also 
obtain the `dual' interpretation of the pressure of the thermal system 
\be
p = -\tilde e_0(\beta)\,.
\ee\
For instance, in a regime where the system behaves in a scale-invariant way,
$s=cT^3$ and $e-3p=0$, the corresponding dual ground-state energy is given by 
\be
\tilde e_0(\beta) = -\frac{c}{4\beta ^4}\,.
\ee

We remark that finite-temperature phase transitions are mapped into 
quantum phase transitions in this interpretation~\cite{sachdev-book}.
The vacuum energy $\tilde e_0(\beta)$ has a non-analyticity 
at a critical value of $\beta$ equal to $1/T_c$.
This non-analyticity is typically due to an avoided level-crossing.

\section{Finite-volume effects on the thermodynamic potentials}
We can exploit the dual interpretation of
the partition function further to study the finite-volume effects on the thermal system. 
Through a chain of relations, we successively relate
the expectation value  of the energy-momentum tensor  on a
$\beta\times L^3$ lattice to the same expectation value  on a
$\beta\times \infty^3$ lattice. This allows us to arrive at a formula 
for the finite-volume correction to the thermodynamic potentials.
Starting with the thermal expectation value of $\theta_{00}$, 
we successively write 
\ba
-\frac{1}{3} \<\theta_{00}\>_{\beta\times L^3} &=& \<\theta_{33}\>_{\beta\times L^3}
= \<\theta_{00}\>_{L\times(L^2\times \beta)} = 
\<\theta_{00}\>_{\infty\times(L^2\times\beta)} + \K_1
\la{eq:chain}
\\
&=& \<\theta_{11}\>_{L\times(\infty\times L\times \beta)}+\K_1
   = \<\theta_{11}\>_{\infty\times(\infty\times L\times \beta)} + \K_1+\K_2
\nn
&=& \<\theta_{00}\>_{\infty\times(\infty\times L\times \beta)} + \K_1+\K_2
= \<\theta_{22}\>_{L\times(\infty\times\infty\times\beta)} + \K_1+ \K_2
\nn
&=& \<\theta_{22}\>_{\infty\times(\infty\times\infty\times\beta)} +\K_1+\K_2+\K_3
= -\frac{1}{3} \<\theta_{00}\>_{\beta\times \infty^3} + \K_1 + \K_2 + \K_3 \,.
\nonumber
\ea
A spectral representation for the $\K_i$ 
is obtained in appendix A, for instance 
\be
\K_1 =  \frac{1}{\beta L^2\,Z\,e^{ \tilde E_0 L}}\sum_{n\geq 1}
\left(\< \tilde \Psi_n|{\txts\int} d^3x\,\hat \theta_{00}(x)| \tilde \Psi_n\>
- \< \tilde \Psi_0|{\txts\int} d^3x\,\hat \theta_{00}(x)| \tilde \Psi_0\>\right)
e^{-( \tilde E_n-  \tilde E_0)L}\,,
\la{eq:K1def}
\ee
where the $|\tilde \Psi_n\>$ and $\tilde E_n$ are the eigenstates and energy 
levels of the $\beta\times L^2$ system.
We can use \eq(\ref{eq:chain}) to produce an 
expression for the finite-volume effects on the entropy density:
\be
s = \frac{4\beta}{3}\<\theta_{00}\>_{\beta\times\infty^3}
= \frac{4\beta}{3}\<\theta_{00}\>_{\beta\times L^3}
  + 4\beta (\K_1+\K_2+\K_3)\,.
\la{eq:s1}
\ee

Following the same steps as for the entropy density, we can 
obtain an expression for the leading finite-volume effects on 
the interaction measure. This case is slightly  simpler, because 
the trace-anomaly operator is a Lorentz scalar:
\ba
\<\theta\>_{\beta\times L^3} &=&
\<\theta\>_{L\times(L^2\times\beta)} = \<\theta\>_{\infty\times(L^2\beta)}+\J_1
\\
&=& \<\theta\>_{L\times(\infty\times L\times\beta)} + \J_1
=  \<\theta\>_{\infty\times(\infty\times L\times\beta)} + \J_1 + \J_2 
\nn
&=&  \<\theta\>_{L\times(\infty\times\infty\times\beta)}+ \J_1 +\J_2
=  \<\theta\>_{\infty\times(\infty\times\infty\times\beta)}+ \J_1 +\J_2+\J_3
\nn
&=& \<\theta\>_{\beta\times\infty^3} + \J_1 + \J_2 + \J_3\,.
\nonumber
\ea
A definition for the $\J_i$ based on the spectral representation
is given in appendix B.
We can apply the same reasoning to the $L^4$ system, sending 
the extent of each direction in turn to infinity. There are then 
four correction terms ($\I_\mu$) instead of three. Therefore we obtain
\be
e-3p = \<\theta\>_{\beta\times \infty^3} -\<\theta\>_{\infty^4}
= \<\theta\>_{\beta\times L^3}-\<\theta\>_{L^4} 
-(\J_1+\J_2+\J_3)+(\I_0+\I_1+\I_2+\I_3)\,.
\la{eq:s2}
\ee

\subsection{The case of finite and discrete screening masses}
The general formulas (\ref{eq:s1}) and (\ref{eq:s2}) can be used 
together with the spectral definition of the $\I_\mu,\J_i,\K_i$
to predict the finite volume effects on the thermodynamic potentials.
In the following, we make a qualitative assumption on the 
spectrum of the theory on a $\beta\times L^2$ hypertorus
with $L\gg \beta$, 
which is in particular relevant to QCD at finite temperature.

We consider the case where the low-lying screening masses are
\emph{discrete} energy levels of the $\beta\times L^2$ system. That is to say,
\be
Lm\equiv L(\tilde E_1-\tilde E_0)\gg 1
\ee
and the next energy levels are simply that same excitation
with non-zero momentum in the `transverse' dimensions of size $L$,
\be
\omega({\bf k_\perp}) = \sqrt{m^2 + {\bf k}_\perp^2},\qquad
{\bf k}_\perp=\frac{2\pi}{L}(n_1,n_2)\,,\quad n_i\in{\bf Z}.
\ee
This lightest screening excitation can potentially have a $\nu$-fold
degeneracy.
The next screening mass is assumed to be separated by a gap from the 
lowest one, $(m_2-m)L\gg 1$. In that situation, the $\K_i$ and $\J_i$ 
can be evaluated in a simple fashion, since they receive contributions
only from one-`particle' states. We use particle in quotes because 
the lowest excitations have only two components of momentum; higher
up in the screening spectrum one expects states with an additional 
energy of order $1/\beta$.

We expect the scenario described above to apply in asymptotically free and conformal 
non-Abelian gauge theories. 
For every relativistic theory, the appropriate regime must be studied in order
to correctly predict the leading finite-volume effects.
As a counterexample to the above scenario,
it is well-known that magnetic fields are not screened in an Abelian plasma.

Since we are interested in the leading finite-volume effects,
in the remainder of this section we write 
equations that hold up to terms of order ${\rm max}(e^{-2mL},e^{-m_2L})$.
In appendix A, we calculate the corrections $\K_{i=1,2,3}$ under these
assumptions and find:
\ba
\K_1 &=&  \frac{\nu\, e^{-mL}}{2\pi \beta L^3}\left[2 + 2mL + \frac{3}{4}m^2L^2
-\frac{mL^2}{4}  \beta \partial_{\beta} m(\beta)\right]
\la{eq:K1}
\\
\K_2=\K_3 &=& -\frac{\nu\,e^{-mL}}{2\pi \beta L^3}\left[1+mL + \quarter m^2L^2
 + mL^2\frac{\beta\partial_\beta m}{4}  \right]
\la{eq:K2}
\ea
Plugging these expressions into \eq(\ref{eq:s1}), 
we obtain our final formula
\be
s -   \frac{4\beta}{3}\<\theta_{00}\>_{\beta\times L^3}
= \frac{m \nu\,e^{-m L}}{2\pi L}
\left[  m(\beta) - 3\beta\partial_\beta m(\beta)\right]
+ \dots
\la{eq:s}
\ee
It shows that the knowledge of the longest spatial correlation 
length $1/m$ as a function of 
temperature $T=1/\beta$ allows one to compute the 
leading finite-volume corrections.

The corrections $\I_\mu$ and $\J_k$ are computed in appendix B
under the same assumptions formulated above.
The zero-temperature volume corrections $\I_\mu$ 
are assumed to be due to $\nu_0$ degenerate states of mass $m_0$. 
We find to leading order 
\ba
\J_1=\J_2=\J_3&=& \frac{m\,e^{-mL}}{2\pi\beta L}
      \left[m+ \beta\partial_\beta m\right]
\la{eq:Ji}
\\
\I_0=\I_1=\I_2=\I_3 &=&  \frac{\nu_0\,m_0^3}{2\pi^2 L} K_1(m_0 L)\,,
\ea
where $K_1$ is the modified Bessel function, $m_0$ is the mass 
gap of the theory on the hypertorus of size $L^3$, and $\nu_0$ is
its degeneracy. For instance, in isospin-symmetric QCD 
there would be $\nu_0=3$ pions.
The final formula for the leading finite-volume effects on 
$e-3p$ follows,
\be
e-3p= \<\theta\>_{\beta\times L^3}-\<\theta\>_{L^4}
-\frac{3m\nu\,e^{-mL}}{2\pi L}\left[m/\beta+ \partial_\beta m\right]
+\frac{2\nu_0 m_0^3}{\pi^2 L}\, K_1(m_0 L)
\,
+ \dots
\la{eq:e-3p}
\ee
Combining \eq(\ref{eq:e-3p}) and (\ref{eq:s}) with
the thermodynamic identity $Ts=e+p$, we find that the 
pressure $p$ is the thermodynamic quantity with the simplest
finite-volume effect:
\be
p = 
-\frac{1}{3}\left(\<T_{kk}\>_{\beta\times L^3} -\<T_{kk}\>_{ L^4}\right)
 + \frac{m^2\nu\,e^{-mL}}{2\pi L\beta}
- \frac{m_0^3\nu_0}{2\pi^2 L}\, K_1(m_0 L)
+ \dots
\la{eq:p}
\ee
When the zero-temperature finite-volume corrections are negligible, 
the pressure computed in finite volume is lower than in the 
thermodynamic limit. Note that in \eq(\ref{eq:p}) 
the pressure $p(L)$ is assumed to be obtained directly 
from the expectation value of $T_{kk}$.
If $p/T^4$ is obtained with the so-called `integral method' 
(see for instance~\cite{Engels:1999tk}), i.e. by integrating
$( e -3p)/T^4$ over temperature 
starting at $T=0$, one should go back to \eq(\ref{eq:e-3p}) 
to compute the finite-volume effects.

\section{Applications}
We give two examples where we expect the formulas derived 
above to apply.
\subsection{Confined phase of SU($N$) gauge theory}
We first consider the pure SU($N$) gauge theory
(see~\cite{Teper:2008yi} for a review of its properties).
Below the deconfining temperature $T_c$, 
the center symmetry associated with the direction of length $\beta$
is unbroken. Correspondingly, the expectation value of the Polyakov loop
vanishes even in the infinite spatial-volume limit.
However, for $N=2$ and 3
the  correlation length associated with the sector of 
non-zero winding number\footnote{For SU(3),
the sector of winding number 2 is equivalent to the sector with winding
number -1, which by charge conjugation 
has the same correlation length as the +1 sector. Therefore there are 
only two sector to discuss (winding 0 and +1).}
becomes very long as the critical temperature
is approached from below.
In fact, in the case of SU(2) gauge theory, it even 
diverges with the 3d Ising exponent~\cite{Svetitsky:1982gs}. 
For SU(3), the correlation length
becomes very long but remains finite. In~\cite{Lucini:2005vg}, it was found that 
\be
m({\txts\frac{1}{T_c}})/ T_c = 0.53(4).
\la{eq:xiTc}
\ee
We can use the reinterpretation of the partition function to 
estimate the leading finite-volume effects on the pressure.
At zero-temperature, the lightest state is the scalar glueball, 
so $\nu_0=1$.
Given its large mass, $M_G/T_c\approx 5.3$~\cite{Meyer:2008tr,Lucini:2003zr}, 
it is not difficult to ensure that the $\I_\mu$ corrections are negligible
by making the box size $L$ large enough. We therefore have 
\be
\frac{p(T_c,L=\infty)}{T_c^4} = \frac{p(T_c,L)}{T_c^4} + \delta \,,
\ee
with 
\be
\delta  = \frac{m^2}{T_c^2} \frac{e^{-(m/T_c)\cdot LT_c}}{2\pi LT_c}
=( 0.0013, ~0.00031) \quad {\rm for}\quad LT_c=(4,~6).
\ee
In fact, the value of $p(T_c)/T_c^4$ is not known precisely, 
but is most likely on the order of 0.02, based 
on available numerical data~\cite{Boyd:1996bx},
or on the pressure exerted by the known 
spectrum of glueballs~\cite{Chen:2005mg,Meyer:2004jc}, 
assuming that they are non-interacting.
Therefore the correction at $LT_c=4$
is not negligible if one aims at a precision of one percent 
on the pressure.

In order to predict  the finite volume correction to the entropy density, 
we need an estimate of the derivative of $m$ with respect to $\beta=1/T$.
For a first idea of the order of magnitude involved, 
we can use the Nambu-Goto formula~\cite{Arvis:1983fp,Luscher:2004ib,Aharony:2009gg}
for $m(\beta)=\sigma_{\rm eff}(\beta)\beta$ with
\be
\frac{\sigma_{\rm eff}(\beta)}{\sigma}= 
\left[1-\frac{2\pi}{3}\frac{1}{\sigma\beta^2}\right]^{\frac{1}{2}}\,,
\la{eq:nago}
\ee
where $\sigma$ is the string tension at $T=0$,
to estimate the derivative of the screening mass with respect to $\beta$.
The finite-volume effect is proportional to 
\be
m(m-3\beta\partial_\beta m) = -2(m^2+\pi\sigma)\,.
\ee
In particular, this quantity is negative, so the 
`effective' entropy density computed in finite volume 
decreases towards the infinite-volume limit (the sign is opposite to 
the volume correction on the pressure). When $m$ becomes small near $T_c$,
the magnitude of the finite volume effect on $s$ is about
$\frac{\sigma\,e^{-mL}}{L}$. We can do a numerical application in the 
SU(3) case. For $LT_c=4$,
using the value of $m(\beta)$ 
given in \eq(\ref{eq:xiTc}) and $T_c/\sqrt{\sigma}\approx0.64$~\cite{Lucini:2003zr}, 
we get $\frac{\sigma\,e^{-mL}}{LT_c^3}\approx 0.08$. 
Since  $s/T_c^3$ itself is about 0.2~\cite{Boyd:1996bx,hm-inprep}
(approaching from the confined phase), this is a large effect indeed.
A box size of $LT_c=9$ is required to reduce this finite-size effect 
to one percent. This corresponds to a length $L$ of about 6fm.

It would be interesting to know in what range of quark masses
this large finite-volume effect persists in full QCD, 
even though the center symmetry responsible for 
the existence of the light mode is badly broken in the presence
of light quarks. Beyond checking that the leading 
correction is numerically small, 
it is also important that the exponent $mL$ be large,
as otherwise corrections that are formally higher order 
can be important.

\subsection{In the deconfined phase}
Let us consider the SU(3) gauge theory in the deconfined phase.
Above $T_c$, we know that the smallest screening mass corresponds
to a state invariant under all symmetries of the theory in a 
 $\beta\times L^2$ box~\cite{Grossman:1993wm,Datta:1998eb} and 
its value is~\cite{Meyer:2008dt}
\be
\beta m(\beta) = 2.62(16),\qquad 2.83(16),\qquad 2.88(10)
\ee
respectively at the temperatures $1.24T_c$, $1.65T_c$ and $2.20T_c$.
Due to these large values of the screening mass, 
the volume correction to the pressure appears to be negligible
already for $LT=4$, $\delta p/T^4 \approx 8\cdot 10^{-6}$.
Recall that the Stefan-Boltzmann pressure is $p_{\rm SB}/T^4= 8\pi^2/45\simeq 1.75$
for $N=3,~N_{\rm f}=0$. 

In the strongly coupled, large-$N$,
${\cal N}=4$ Super-Yang-Mills theory the screening masses have been calculated 
by AdS/CFT methods~\cite{Brower:2000rp,Bak:2007fk}. They turn out to be significantly larger than
in QCD, so that finite-volume effects would be even smaller for the 
same value of $LT$.

At asymptotic temperatures in both QCD and SU(3) gauge theory, 
the smallest screening mass corresponds to the $A_1^{++}$ state
of three-dimensional gauge theory, 
with a mass $m/g_3^2\approx 2.40$~\cite{Hart:2000ha,Teper:1998te},
and $g_3^2=g^2(T)T$ to leading order.
When the coupling reaches the value it takes on the Z pole, 
$\alpha_s\approx 0.11$, this means that $\delta p/T^4 \approx 8\cdot 10^{-7}$
for $LT=4$. We conclude that the aspect ratio $LT$ has to be increased only very slowly 
with temperature in order to accomodate the magnetic screening length $1/m\sim 1/g^2T$.

\section{Concluding remarks}
We have derived a simple way to calculate the finite-volume effects affecting the 
energy density and pressure of a relativistic theory at zero chemical potential
in terms of the spectrum of the same theory defined on a 
spatial hypercube with two large cycles and one of length $\beta=1/T$.
When that spectrum is discrete, the leading finite-volume effects can be 
calculated completely in terms of the mass gap. It is almost obvious that 
the finite volume effects should be of order $e^{-mL}$, but we have 
shown that the prefactor is also entirely determined by the screening spectrum and its 
temperature dependence. This is because the diagonal matrix elements 
of the energy-momentum tensor are themselves given in terms of that spectrum
(see \eq(\ref{eq:sr}--\ref{eq:th33})).

It is hoped that \eq(\ref{eq:s}) and (\ref{eq:e-3p}) will be useful in 
controlling  the finite-volume effects in lattice QCD thermodynamics calculations.
If the screening mass gap is known and a finite-volume study shows that the 
finite-size effects are well described by the formula, one can use it to 
correct for these effects. If there are several screening masses below the 
threshold of $2m(\beta)$, those states will contribute terms 
to the finite-size effects similar to the lightest one. To include the effects 
of screening states above $2m(\beta)$, one presumably needs to know the 
scattering length of these `particles'. In QCD at low temperatures, 
explicit calculations using chiral perturbation theory are then likely to be 
predictive, since information on the scattering lengths of pions is available.

It is clear that the method followed here is not specific to four dimensions.
It also applies for instance  to three-dimensional gauge theories. 
The main difference is that the transverse momentum of the lightest screening
state only has one component, so that momentum integrals as in \eq(\ref{eq:K1mom-integ}) 
become one-dimensional.

In SU($N$) gauge theories, it is interesting to note the dependence 
of the finite volume effects of energy density and pressure 
on the number of colors $N$.
We showed that the asymptotic finite volume effect is driven 
by a unique color singlet state and that its contribution is therefore
O($N^0$). In the deconfined phase, the thermodynamic potentials are O($N^2$),
and the relative size of finite-volume effects is thus $1/N^2$ suppressed.
This conclusion
remains qualitatively valid in the presence of quarks, since their main effect is to 
add a contribution of order $NN_{\rm f}$ to the thermodynamic potentials.

In the confined phase on the other hand, the thermodynamic potentials
are O($N^0$), so there is no parametric suppression of the volume effect there.
Since on the lattice the entropy density is simply computed as the difference 
between the $1\times1$ electric and magnetic Wilson loops (`plaquettes'), 
this might seem to contradict
the statement that finite volume effects on expectation values of Wilson loops vanish 
in the large-$N$ limit as long as the center symmetries remain intact. However 
the statement of volume-independence only applies to the  O$(N^2)$ contribution 
to the plaquette (see section 2.3 of~\cite{Kovtun:2007py} for a clear discussion); 
the latter is divergent in the continuum limit and cancels in the difference 
of electric and magnetic plaquettes. Therefore there is no contradiction between
our results and the large-$N$ volume-independence arguments.

\acknowledgments{
This work was supported in part by
funds provided by the U.S. Department of Energy 
under cooperative research agreement DE-FG02-94ER40818.
}
\appendix
\section{Calculation of $\K_{1,2,3}$}
In this appendix, we compute the quantities $\K_i$ that 
are the finite-time extent corrections to certain expectation values 
of $\theta_{\mu\nu}$. 
Let us start with $\K_1$, it is 
the difference between the expectation value of $ \<\theta_{00}\>$
on a $L\times(L^2\times \beta)$ lattice and on an
$\infty\times(L^2\times \beta)$ lattice. 
(In this appendix, we drop the $\tilde{}$ on the energies and states
of the $\beta\times L^2$ system, and we set
the degeneracy $\nu$ of the energy level $E_1$ to one,  since the more general 
result is simply obtained by multiplying $\K_1$ by $\nu$.)
\ba
\beta L^2 \<\theta_{00}\>_{L\times(L^2\times \beta)} &=&
\frac{1}{Z(L,L,L,\beta)} \sum_{n} e^{-E_n L}
\< \Psi_n|\int d^3x\,\hat \theta_{00}(x)| \Psi_n\>
\\
&=& \< \Psi_0|\int d^3x\,\hat \theta_{00}(x)| \Psi_0\>
\nn
&+& \frac{1}{Ze^{ E_0 L}}\sum_{n\geq 1}
\left(\< \Psi_n|{\txts\int} d^3x\,\hat \theta_{00}(x)| \Psi_n\>
- \< \Psi_0|{\txts\int} d^3x\,\hat \theta_{00}(x)| \Psi_0\>\right)
e^{-( E_n-  E_0)L}\,.
\nonumber
\ea
We  now observe that 
\be
\< \Psi_0|\int d^3x\,\hat \theta_{00}(x)| \Psi_0\> 
= \beta L^2\<\theta_{00}\>_{\infty\times(L^2\times\beta)}\,.
\ee
We therefore identify $\K_1$ as (see \eq(\ref{eq:chain}))
\be
\K_1 =  \frac{1}{\beta L^2\,Z\,e^{ E_0 L}}\sum_{n\geq 1}
\left(\< \Psi_n|\int d^3x\,\hat \theta_{00}(x)| \Psi_n\>
- \< \Psi_0|\int d^3x\,\hat \theta_{00}(x)| \Psi_0\>\right)
e^{-( E_n-  E_0)L}\,.
\ee
We now assume that the lowest-lying excited states are `one-particle' excitations
with arbitrary momentum in the two directions of length $L$.
Let us therefore call $  m=   E_1-  E_0$ the energy of the
first excited state, since it is at rest.
By rotation symmetry among the three dimensions that are not of length $\beta$,
the dispersion relation must be relativistic and we define
\be
  \omega({\bf k}) \equiv \sqrt{  m^2+{\bf k}^2}\,.
\ee
Here ${\bf k}$ is a two-component vector. 
The partition function is for instance  given by
\be
Ze^{  E_0 L} = \sum_{n} e^{-(  E_n-  E_0) L}
= 1 +  \sum_{\bf k} 
 e^{-  \omega({\bf k}) L} + {\rm O}(e^{-2mL})\,.
\ee
but to the order we are working at, we can use $Ze^{  E_0 L} =1$.
We are thus lead to the expression
\be
\K_1 = \frac{1}{\beta L^2}  \sum_{\bf k}
e^{- \omega({\bf k})L}
\left(\<  \Psi_1({\bf k})|{\txts\int} d^3x\,\hat \theta_{00}(x)|  \Psi_1({\bf k})\>
-\<  \Psi_0|{\txts\int} d^3x\,\hat \theta_{00}(x)|  \Psi_0\> \right)
+ \dots
\la{eq:K1mom-integ}
\ee
Using \eq\ref{eq:sr}, we obtain
\[
\<  \Psi_1({\bf k})|{\txts\int} d^3x\,\hat \theta_{00}(x)|  \Psi_1({\bf k})\>
- \<  \Psi_0|{\txts\int} d^3x\,\hat \theta_{00}(x)|  \Psi_0\>
=    \omega({\bf k}) - \frac{1}{4} \frac{  m }{ \omega({\bf k})}
 \frac{\partial}{\partial \beta } (  m  \beta )\,.
\]
Using the Poisson summation formula and performing the integral, one finds
\ba
\K_1 &=& \frac{1}{2\pi\beta}\sum_{\bf n}
\frac{e^{-m\sqrt{{\bf y}_n^2+L^2}}}{\sqrt{{\bf y}_n^2+L^2}}
\Big[-\quarter m\partial_\beta(m\beta)
\\ &&
+\frac{L^4m^2- {\bf y}_n^2(1+m\sqrt{{\bf y}_n^2+L^2})+L^2(2+m^2{\bf y}_n^2+2m\sqrt{{\bf y}^2+L^2})}
{({\bf y}_n^2+L^2)^2}\Big]\,.
\nonumber
\ea
Here ${\bf y}_n\equiv L {\bf n}$ and ${\bf n}\in {\bf Z}^2$. 
It is now obvious that the terms with ${\bf n}\neq 0$ are subleading 
and can be dropped, hence
\be
 \K_1 =\frac{e^{-  m  L}}{2\pi \beta L^3}\left[2 + 2  m  L +   m ^2 L^2
-\quarter   m   L^2 \frac{\partial}{\partial \beta } (  m  \beta )\right]
+{\rm O}(e^{-2mL})\,.
\ee
Since we are assuming that $mL\gg 1$,  we can always replace
the momentum sum for a direction of size $L$ by an integral,
$\frac{1}{L}\sum_k\to\int \frac{dk}{2\pi}$. The corrections to this 
are suppressed by $e^{-mL}$, as just shown.
It is therefore clear from their definitions that $\K_2=\K_3$ to this accuracy.
These expressions are calculated in the same fashion as $\K_1$, using 
this time \eq(\ref{eq:th33}).
The contribution of the $|\Psi_1({\bf k})\>$ states reads
\be
\K_2 = -\frac{m}{4\beta}\int \frac{d^2{\bf k}}{(2\pi)^2} e^{-\omega({\bf k})L}
\left[\frac{2\omega({\bf k})}{m} +  \frac{\beta\partial_\beta m - m}{\omega({\bf k})}\right]
+\dots,
\ee
which leads to \eq(\ref{eq:K2}).

\section{Calculation of $\J_{1,2,3}$}
The derivation follows closely that of appendix A. 
The spectral representation for $\J_1$ is 
\be
\J_1 = \frac{1}{\beta L^2\,Z\,e^{E_0L}}
\sum_{n\geq 1} 
\left(\<\Psi_n|{\txts\int} d^3x\,\hat \theta(x)|  \Psi_n)\>
- \<  \Psi_0|{\txts\int} d^3x\,\hat \theta(x)|  \Psi_0\>\right)
e^{-(E_n-E_0)L}\,.
\la{eq:J1def}
\ee
We use \eq(\ref{eq:sr2}) to reach the expression
\be
\J_1 = \frac{m}{\beta}\left(m+\beta\partial_\beta m\right)
\int \frac{d^2{\bf k}}{(2\pi)^2}
\frac{e^{-\omega({\bf k})L} }{\omega({\bf k})}
\ee
Performing the momentum integral leads to \eq(\ref{eq:Ji}).
It is clear that the $\J_i$ only differ by the size of the 
spatial dimensions transverse to the dimension of size $\beta$
(they are either of size $L$ or infinite).
Since we assume $mL\gg1$, this difference is subleading, as can 
easily be seen by using the Poisson summation formula.

As for $\I_0$, the same definition as \eq(\ref{eq:J1def}) holds, except 
that the states live on an $L\times L\times L$ hypercube.
Therefore we get (for a degeneracy $\nu_0$ of 1 and setting 
$\omega({\bf k}) = \sqrt{{\bf k}^2+m_0^2}$, where $m_0$ is the mass gap)
\be
\I_0 = m_0^2\int \frac{d^3{\bf k}}{(2\pi)^3}
           \frac{e^{-\omega({\bf k})L} }{\omega({\bf k})}
= \frac{m_0^3}{2\pi^2 L} K_1(m_0L)\,.
\ee
For bosonic degrees of freedom, the same calculation applies to 
the $\I_k$, neglecting terms of order $e^{-2m_0L}$. Therefore all $\I_\mu$
are equal. Indeed all directions are truly symmetric if the boundary conditions
are periodic. 
For fermions, strictly speaking the boundary condition has to be antiperiodic
in all directions for the same to apply. Indeed it is for antiperiodic 
boundary conditions that the path integral computes the trace over states.
\bibliographystyle{JHEP}
\bibliography{/afs/lns.mit.edu/user/meyerh/BIBLIO/viscobib}

\end{document}